\def\xis{{\xi_{\ast}}}

\documentstyle[11pt]{article}

\def\appendix{\par
 \setcounter{section}{0}
 \setcounter{subsection}{0}
 \def\thesection{Appendix \Alph{section}}
 \def\theequation{\Alph{section}.\arabic{equation}}
 \setcounter{equation}{0}}

\begin{document}
\begin{flushright}
DO-TH 93-11 \\
(August 1993)
\end{flushright}
\centerline{\Large \bf CP-Violation in Bosonic Sector \\
                       of SM with two Higgs Doublets}
\vspace{1.2cm}
\begin{center}
{{\bf G.~Cveti\v c} \\
Inst.~f\"ur Physik, Universit\"at Dortmund, 44221 Dortmund, Germany}
\end{center}
\vspace{1.2cm}
\centerline{\bf ABSTRACT}

We investigate CP-violation effects in the bosonic sector of the Standard
Model (SM) with
two Higgs doublets. First we calculate the mass eigenstates of the physical
neutral Higgses for small but nonzero CP-violation parameter $\xi_{\ast}$,
and then a ``forward-backward'' asymmetry ${\cal A}_{fb}$ for the decay
$H \to W^+W^-Z$ that would be a signal of CP-violation. Although the
effects are in general small (${\cal A}_{fb}
= \Gamma_{fb}/\Gamma \sim {\cal O}(10^{-3})$), ${\cal A}_{fb}$
turns out to be a rather clean signal of CP-violation, since neither the
CP-conserving final state interactions nor the direct production
background events contribute to $\Gamma_{fb}$.
The CKM-type CP-violation effects that could in principle also
contribute to ${\cal A}_{fb}$ are negligible.
The nonzero ${\cal A}_{fb}$ could possibly be detected at some
later stage at the LHC or SSC.

\newpage

The Standard Model (SM) with two Higgs doublets and CP-violation in the
bosonic sector
has recently drawn a lot of attention~\cite{sw}, especially because it
could give an explanation of the baryogenesis~\cite{ckn,tz}, unlike the
CP-violation originating from the Cabibbo-Kobayashi-Maskawa (CKM) matrix
of the SM.

The most general gauge invariant potential for
two Higgs doublets $\Phi_i$ ($i=1,2$)
with $Y=1$ which induces only suppressed flavor-changing neutral currents
(FCNC's) is~\cite{ghkd}
\begin{eqnarray}
V(\Phi_1,\Phi_2) & = & \lambda_1 (\Phi_1^{\dagger}\Phi_1 - v_1^2)^2 +
  \lambda_2 (\Phi_2^{\dagger}\Phi_2 - v_2^2)^2
 \nonumber\\
 & & + \lambda_3 [(\Phi_1^{\dagger}\Phi_1 - v_1^2)
   + (\Phi_2^{\dagger}\Phi_2 - v_2^2)]^2 +
 \lambda_4[(\Phi_1^{\dagger}\Phi_1)(\Phi_2^{\dagger}\Phi_2) -
    (\Phi_1^{\dagger}\Phi_2)(\Phi_2^{\dagger}\Phi_1)]
\nonumber\\
 & &  + \lambda_5 [Re(\Phi_1^{\dagger}\Phi_2) - v_1 v_2 \cos \xi ]^2
  + \lambda_6 [Im(\Phi_1^{\dagger}\Phi_2) - v_1 v_2 \sin \xi ]^2  \ .
\end{eqnarray}
The potential spontaneously breaks $SU(2)_L \times U(1)_Y$ down to
$U(1)_{EM}$. The fact that the discrete symmetry $\Phi_1 \mapsto -\Phi_1$
is only softly violated (by terms of dimension two) guarantees that the
FCNC's are not too large. The six parameters $\lambda_i$ are in general
of the order $(M_{scalar}/v)^2 \sim {\cal O} (1)$. The minimum of the
potential is at
\begin{equation}
\langle \Phi_1 \rangle_o = { 0 \choose v_1} \ , \qquad
\langle \Phi_2 \rangle_o = e^{i \xi} { 0 \choose v_2 } \ ,
 \qquad (v_1^2+v_2^2 = \frac{1}{2} v^2 = \frac{1}{2} 246^2 GeV^2) \ .
\end{equation}
For $\xi_{\ast} \ (=(\lambda_5 - \lambda_6 ) \xi ) = 0$, we have no
CP-violation, and the neutral physical scalars $H^o_+$,$h^o_+$,$A^o_-$ have
well-defined CP (equal to $+1,+1,-1$, respectively). These scalars and
their masses are known (we use the notations of~\cite{ghkd})
\begin{eqnarray}
\left( \begin{array}{c}
        H^o_+ \\
        h^o_+
       \end{array} \right)
& = &  \sqrt{2}
\left( \begin{array}{rc}
        \cos \alpha & \sin \alpha \\
      - \sin \alpha & \cos \alpha
       \end{array} \right)
\left( \begin{array}{c}
        Re\Phi^o_1-v_1 \\
        Re\Phi^o_2-v_2
       \end{array} \right) \ ,
\nonumber\\
A^o_- & = & \sqrt{2} ( - Im \Phi^o_1 \sin \beta + Im \Phi^o_2 \cos \beta ) \ ,
\end{eqnarray}
where
\begin{displaymath}
\beta = \arctan (\frac{v_2}{v_1}) \ , \qquad
\alpha = \frac{1}{2} \arctan \frac{ 2 {\cal M}_{12} }
{ {\cal M}_{11} - {\cal M}_{22} } \ ,
\qquad sgn(\sin 2 \alpha ) = sgn( {\cal M}_{12} ) \ ,
\end{displaymath}
\begin{displaymath}
{\cal M}  =
\left( \begin{array}{cc}
       4 v_1^2 (\lambda_1 + \lambda_3 ) + v_2^2 \lambda_5 &
       ( 4 \lambda_3 + \lambda_5 ) v_1 v_2  \\
       ( 4 \lambda_3 + \lambda_5 ) v_1 v_2    &
       4 v_2^2 (\lambda_2 + \lambda_3 ) + v_1^2 \lambda_5
       \end{array} \right) \ ,
\end{displaymath}
\begin{equation}
{M_{H^o}^2 \choose M_{h^o}^2} = \frac{1}{2} [ {\cal M}_{11} + {\cal M}_{22}
\pm \sqrt{({\cal M}_{11} - {\cal M}_{22})^2 + 4 {\cal M}_{12}^2} ] \ ,
\qquad M_{A^o} = \frac{1}{2} \lambda_6 v^2 \ .
\end{equation}
On the other hand, in the case of $\xi_{\ast} = (\lambda_5 - \lambda_6) \xi
\not= 0$ we do have CP-violation. It is possible to find the physical scalar
mass eigenstates in this case, if we make the expansion in powers of $\xi$
in the potential (1). Denoting the mass eigenstates ($H^o_+$, $ \ h^o_+$,
$ \ A^o_-$)
of the $\xi_{\ast}=0$ case as $v_{(1)}(0)$,$ \ v_{(2)}(0)$,$ \ v_{(3)}(0)$,
respectively, we obtain the three mass eigenstates $v_{(j)}(\xi_{\ast})$
after some lengthy algebra:
\begin{equation}
v_{(j)}(\xi_{\ast}) = U_{jk}(\xi_{\ast}) v_{(k)}(0) \ ,
\end{equation}
where $U(\xis)$ is a $3 \times 3$ orthogonal matrix:
\begin{eqnarray}
U_{11}(\xis) & = & 1-\frac{1}{2} \xis^2 \frac{\sin^2 \delta}{(x_3-x_1)^2}
 + {\cal O}((\lambda_5-\lambda_6)^2\xi^4)
\ , \nonumber\\
U_{22}(\xis) & = & 1 - \frac{1}{2} \xis^2 \frac{\cos^2 \delta}{(x_3-x_2)^2}
 + {\cal O}((\lambda_5-\lambda_6)^2\xi^4)
\ , \nonumber\\
U_{33}(\xis) & = & 1 - \frac{1}{2} \xis^2 [\frac{\sin^2 \delta}{(x_3-x_1)^2}
 + \frac{\cos^2 \delta}{(x_3-x_2)^2}]+{\cal O}((\lambda_5-\lambda_6)^2\xi^4)
\ , \nonumber\\
U_{12}(\xis) & = & \xi^2 (\lambda_5-\lambda_6)
\frac{(\lambda_5-x_1) \sin (2 \delta)}
{ 2 (x_1-x_2) (x_1-x_3) }+ {\cal O}((\lambda_5-\lambda_6)\xi^3)
\ , \nonumber\\
U_{21}(\xis) & = & \xi^2 (\lambda_5-\lambda_6)
\frac{(\lambda_5-x_2) \sin (2 \delta)}
{2 (x_2-x_1) (x_2-x_3)}+ {\cal O}((\lambda_5-\lambda_6)\xi^3)
\ , \nonumber\\
U_{31}(\xis) & = & \xis \frac{\sin \delta}{(x_1-x_3)} +
{\cal O}((\lambda_5-\lambda_6)\xi^3)  =  - U_{13}(\xis)
\ , \nonumber\\
U_{32}(\xis) & = & \xis \frac{\cos \delta}{(x_2-x_3)} +
{\cal O}((\lambda_5-\lambda_6)\xi^3)  =  - U_{23}(\xis) \ ,
\end{eqnarray}
where
\begin{equation}
\xis=(\lambda_5-\lambda_6)\xi \ , \qquad \delta = \beta + \alpha \ ,
\qquad x_j = 2 M_j^2(0)/v^2  \qquad (j=1,2,3) \ .
\end{equation}
Here, $M_j(0)$ ($j=1,2,3$) are masses of $v_{(j)}(0)$ (of eq.~(4)).
The masses of $M_j(\xis)$ of $v_{(j)}(\xis)$ differ from those of $\xis=0$
case only slightly (for $\xi \ll 1$)
\begin{equation}
M_j^2(\xis) = M_j^2(0) + \xi^2 (\lambda_5-\lambda_6) \frac{v^2}{2}Y_j
+ {\cal O}((\lambda_5-\lambda_6)\xi^4) \ ,
\end{equation}
where
\begin{eqnarray}
Y_1 & = & \frac{(x_1-\lambda_5)}{(\lambda_6-x_1)} \sin^2 \delta \ ,
\qquad Y_2  =  \frac{(x_2-\lambda_5)}{(\lambda_6-x_2)} \cos^2 \delta \ ,
\nonumber\\
Y_3 & = & 1 - (\lambda_5-\lambda_6)\left[ \frac{\cos^2 \delta}{x_2-\lambda_6}
+ \frac{\sin^2 \delta}{x_1-\lambda_6} \right] \ .
\end{eqnarray}
The quantities $\delta$ and $x_j$ are dimensionless, in general of order
${\cal O}(1)$. Note that the charged scalar sector remains unaffected by
the introduction of $\xis \not= 0$.

Equation (5) tells us that the mass eigenstates of the neutral physical
scalars are in general linear combinations of $CP=+1$ and $CP=-1$
components. This feature of CP-violation could be tested experimentally
by looking at the decays of the (heavy) Higgs $H^o$ ($=v_{(1)}(\xis)$)
to $W^+W^-Z$, as proposed within a more general context in ref.~\cite{cnp}.
As argued there, in the unitary gauge, the decay amplitude $T_+$
mediated by $W^{\ast \pm}$
and $Z^{\ast}$ exchange (Figs.~1a,b) would yield final states with
$CP=+1$ (just like in the minimal SM), while the decay amplitude $T_-$
mediated by (neutral) physical scalars (Fig.~2) would yield final
states with $CP=-1$ \footnote{The latter amplitude is zero if $\xis = 0$,
as expected.}. On the other hand, the two-body decays $H^o \to W^+W^-, ZZ$
at the tree level would not test the mixed CP-structure of $H^o$, because
the final state is an S-wave due to the coupling without derivatives
($L_{fin}=0 \Rightarrow S_{fin}=0 \Rightarrow CP(W^+W^-)=(-1)^{S_{fin}}
= +1 = CP(ZZ)$).

For the decay $H^o \to W^+W^-Z$,
we can construct the following experimentally relevant ``forward-backward''
asymmetry width parameter $\Gamma_{fb}$ which would be a signal of this
CP-violation
\begin{displaymath}
\Gamma_{fb}(H^o \to W^+W^-Z) =
\end{displaymath}
\begin{equation}
\left[ \int_0^{+1} d( \cos \theta)
\frac{d \Gamma (H^o \to W^+W^-Z)}{d( \cos \theta)} -
\int_{-1}^{0} d( \cos \theta)
\frac{d \Gamma (H^o \to W^+W^-Z)}{d( \cos \theta)} \right] \ ,
\end{equation}
where $\theta$ is the angle between $ \vec p_{W^+}$ and $- \vec p_Z$
in the frame CMS($W^+W^-$) (Fig.~3), and the sum over the helicities of the
final particles is implicitly assumed
\footnote{Note that $\Gamma_{fb}$
is in principle obtained by measuring the corresponding ``forward-backward''
difference $N_{fb}$ of the number of these decays: $\Gamma_{fb} = N_{fb}
\frac{\Gamma(H^o)}{L \sigma}$, where $\Gamma(H^o)$ the total decay width,
$L$ the integrated luminosity and $\sigma$ the production cross section for the
(heavy)
Higgs.}.

Note that $\theta \mapsto \pi - \theta$ under the CP-transformation of the
final state $W^+W^-Z$, and hence $\Gamma_{fb} \mapsto - \Gamma_{fb}$ if
$H^o$ were a pure $CP=+1$ or $CP=-1$ state. Therefore, in the case of
CP-conservation we must have $\Gamma_{fb} = 0$. Hence, $\Gamma_{fb}
\neq 0$ is a signal of CP-violation. We can show in general, using
the formalism of partial wave expansions of decay amplitudes~\cite{w}, that
$\Gamma_{fb}$ is an expression proportional
to the interference terms of the $CP= \pm 1$ decay amplitudes
\begin{equation}
\Gamma_{fb}(H^o \to W^+W^-Z) \propto \int d(config.space)
( T^{\ast}_-T_+ + T_-T^{\ast}_+ ) \ sgn(\cos \theta) \ .
\end{equation}
In the specific model at hand, we can check this by explicit calculation.
The tree level transition amplitudes $T_{\pm}$
\footnote{Strictly speaking these are not tree level amplitudes, because
the dominant final state interactions in the propagator are taken into
account by nonzero widths of the mediating bosons. The amplitudes can be
calculated in any $R_{\zeta}$-gauge, do not depend on the logitudinal parts
of the propagators and have no $\zeta$-dependence.}
in this case turn out to be
\begin{equation}
T_{\pm} = \epsilon^{\mu_1}(p_1 h_1) \epsilon^{\mu_2}(p_2 h_2)
 \epsilon^{\mu_3}(p_3 h_3) {\cal F}^{(\pm)}_{\mu_1 \mu_2 \mu_3}
(p_1,p_2,p_3) \ ,
\end{equation}
where $p_j$ and $h_j$ ($j=1,2,3$) denote momenta and helicities of
$W^+$, $W^-$ and $Z$, respectively, and
\begin{displaymath}
{\cal F}^{(+)}  =  {\cal F}^{(+,1)} + {\cal F}^{(+,2)}
 + {\cal F}^{(+,3)} \ ,
\end{displaymath}

\begin{eqnarray}
{\cal F}^{(+,1)}_{\mu_1 \mu_2 \mu_3}(p_1p_2p_3)  & = &
\frac{g^2 \cos^2 \theta_{w} M_Z (U_{11} \cos \eta + U_{12} \sin \eta)}
{[ (p_2 + p_3)^2 - M_W^2 + i \Gamma_W M_W ]} \times
\nonumber\\
\lefteqn{ \left[ 2 (p_{2 \mu_3} g_{\mu_1 \mu_2} - p_{3 \mu_2} g_{\mu_1 \mu_3})
 - (p_2 - p_3)_{\mu_1} g_{\mu_2 \mu_3} - \frac{ \sin^2 \theta_w}
{\cos^2 \theta_w} (p_1 + p_2 + p_3)_{\mu_1} g_{\mu_2 \mu_3} \right] \ , }
\nonumber\\
{\cal F}^{(+,2)}_{\mu_1 \mu_2 \mu_3}(p_1p_2p_3) & = &
- {\cal F}^{(+,1)}_{\mu_2 \mu_1 \mu_3}(p_2p_1p_3) \ ,
\nonumber\\
{\cal F}^{(+,3)}_{\mu_1 \mu_2 \mu_3}(p_1p_2p_3) & = &
\frac{g^2 M_Z (U_{11} \cos \eta + U_{12} \sin \eta)}
{[ (p_1 + p_2)^2 - M_Z^2 + i \Gamma_Z M_Z ]} \times
\nonumber\\
& & \left[ 2 (p_{1 \mu_2} g_{\mu_1 \mu_3} - p_{2 \mu_1} g_{\mu_2 \mu_3})
 - (p_1 - p_2)_{\mu_3} g_{\mu_1 \mu_2} \right] \ ,
\end{eqnarray}

\begin{equation}
{\cal F}^{(-)}_{\mu_1 \mu_2 \mu_3} (p_1 p_2 p_3)  =
\frac{ i g^2 M_W }{ \cos \theta_w } g_{\mu_1 \mu_2} (p_1 + p_2 + p_3)_{\mu_3}
{\cal A}(p_1 \cdot p_2) \ ,
\end{equation}
\begin{displaymath}
{\cal A}( p_1 \cdot p_2 ) = \sum_{j=1}^{3} A_j [(p_1+p_2)^2 - M_j^2 +
i \Gamma_j M_j ]^{-1} \ ,
\end{displaymath}
\begin{eqnarray}
A_j & = & U_{13} [ \cos \eta \sin \eta (U_{j1}^2 - U_{j2}^2)
- \cos (2 \eta ) U_{j1}U_{j2} ] +
\nonumber\\
& &  U_{12} [ \cos^2 \eta U_{j1} U_{j2} + \cos \eta \sin \eta U_{j2} U_{j3} ]
-  U_{11} [ \cos \eta \sin \eta U_{j1} U_{j3} + \sin^2 \eta U_{j2} U_{j3} ]
 \ .
\end{eqnarray}

$M_j$ are the masses of the three physical scalars ($M_j = M_j(\xis)
 \simeq M_j(0)$), $\Gamma_j$ are the corresponding widths, and $\eta =
(\beta - \alpha)$. In this particular
case, we explicitly see that $\mid T_+ \mid^2$ and $\mid T_- \mid^2$ are
symmetric under $p_1 \leftrightarrow p_2$, while $(T^{\ast}_+ T_- +
T_+T^{\ast}_-)$ is antisymmetric (summation over the final helicities $h_j$
is always assumed). Therefore, $\mid T_+ \mid^2$ and $\mid T_- \mid^2$
contribute to $\Gamma$ and not to $\Gamma_{fb}$, while $(T^{\ast}_+ T_- +
T_+T^{\ast}_-)$ contributes to $\Gamma_{fb}$ and not to $\Gamma$ (for the
decay $H^o \to W^+W^-Z$). Hence, we see explicitly that relation (11)
holds in the specific discussed case.

In the further calculation, we will assume that $M_1$ ($=M_{H^o}$)
$ > (2 M_W + M_Z)$ and $M_2, \ M_3 < 2 M_W$, and that $\Gamma_2$ and
$\Gamma_3$ are consequently negligible ($\Gamma_2, \Gamma_3 \ll \Gamma_Z
\simeq 2.5 GeV$). The asymmetry signal $\Gamma_{fb}$ would then be
proportional to $\Gamma_W$ and $\Gamma_Z$. Furthermore, we will assume
$\xi < 1$ and $\xis = (\lambda_5 - \lambda_6) \xi < 1$, in order to use
the expressions (6). Then it follows
\begin{displaymath}
{\cal A}(p_1 \cdot p_2)  \simeq   \frac{(A_{l=2})(M_2^2-M_3^2)}
 {[(p_1+p_2)^2-M_2^2] [(p_1+p_2)^2-M_3^2]}
 \left[1 + {\cal O} ( \xi \xis ) \right] \ ,
\end{displaymath}

\begin{equation}
\Gamma_{fb}(H^o \to W^+W^-Z)  \simeq  \Delta \times (\cos \delta
  \cos \eta \sin^2 \eta) \xis
 \left[ 1 + {\cal O} ( \xi \xis ) \right] \ .
\end{equation}
The width $\Delta$ in the above formula
\footnote{Strictly speaking, $\Gamma_{fb} \simeq \Delta \times
(\cos \delta \cos \eta \sin^2 \eta)  \left[ 1 +
\frac{M_2^2-M_3^2}{M_1^2-M_3^2} \tan \delta \cot \eta \right] \xis
 + {\cal O} (\xi \xis^2 \Delta )$, \\
($\delta = \beta + \alpha$, $\eta = \beta - \alpha$) , which reduces
to the above form for $M_2^2, M_3^2 \ll M_1^2$, or for $M_2 = M_3$.}
is
\begin{equation}
\Delta = \frac{M^2_W v^2 g^4}{M_{H^o}^3 2^8 \pi^3 \cos^2 \theta_w}
 \times I( \{ \frac{M_j}{M_W} \} ) \simeq \frac{2700 GeV^4}{M_{H^o}^3}
 \times I \ ,
\end{equation}
where $I ( \{ M_j/M_W \} )$ is a specific
``forward-backward'' asymmetry integral	on the corresponding Dalitz
plot, containing suppression factors $\Gamma_W/M_W$ and $\Gamma_Z/M_Z$.
Numerical calculations yield for $M_1$ ($=M_{H^o}$) $= 400-800 GeV$ and
$M_2,M_3 = 0-100 GeV$:
$ \ 0.23 \stackrel{<}{\sim} I \stackrel{<}{\sim} 90$.
The corresponding values for $\Delta$ are
given in Table 1. If assuming $0.2 < \xis (< 1)$ and $\cos \delta
\cos \eta \sin^2 \eta \stackrel{>}{\sim} 0.5$, (16) gives
\begin{displaymath}
  \Gamma_{fb}(H^o \to W^+W^-Z) \stackrel{>}{\sim} (0.1)\Delta \ .
\end{displaymath}
We may also construct the dimensionless asymmetry parameter
\begin{equation}
{\cal A}_{fb}	= \frac{\Gamma_{fb}(H^o \to W^+W^-Z)}
{\Gamma(H^o \to W^+W^-Z)} = \frac{N_{fb}}{N} \simeq
\rho \frac{\cos \delta \sin^2 \eta}{\cos \eta} \xis
[ 1 + {\cal O} (\xi \xis ) ] \ .
\end{equation}
The dimensionless parameter $\rho$ is small (${\cal O} (10^{-3})$), due to
the suppression factors $\Gamma_W/M_W$ and $\Gamma_Z/M_Z$, and its values
are also included in Table 1.

Here we have to mention that the relation (18) is not valid in the limiting
case of $\cos \eta \to 0$, because in such a case $\Gamma(H^o \to
W^+W^-Z) \simeq 0 + {\cal O} ( \xis^2 )$. In such
a case, ${\cal A}_{fb}$ could be large ($ \stackrel{<}{\sim} {\cal O}(1)$).

For completeness, we also write the decay width $\Gamma (H^o \to
W^+W^-Z)$ in the theory discussed here
\begin{displaymath}
\Gamma(H^o \to W^+W^-Z)  =  \Gamma_+ + \Gamma_-   ,
\end{displaymath}
where
\begin{eqnarray}
\Gamma_+ & = & \Gamma (H^o \to (W^+W^-Z)_{CP=+1}) =
(U_{11} \cos \eta + U_{12} \sin \eta)^2  \ \Gamma^{MSM} (H^o \to W^+W^-Z)
\nonumber\\
& \simeq & \cos^2 \eta \ \Gamma^{MSM}(H^o \to W^+W^-Z) [ 1
 + {\cal O} ( \xi \xis )] \ ,
\end{eqnarray}
\begin{equation}
\Gamma_- = \Gamma(H^o \to (W^+W^-Z)_{CP=-1} ) \simeq
(\cos^2 \delta \sin^4 \eta)  \xis^2 G
\left[ 1 + {\cal O} ( \xi^2 ) \right] \ ,
\end{equation}
where the widths $G$, as well as $\Gamma^{MSM}(H^o \to W^+W^-Z)$ of the
minimal SM, are given in Table 2 for various values of the scalar masses
\footnote{Note that the angle $\eta = \beta - \alpha$ may be obtained
from experiments measuring the $W^+-W^-$-Higgs couplings. The angle
$\beta$ may be restricted indirectly by experiments whose results depend
on the ratio of the vacuum expectation values, within the considered
theory.}.
Note that $\Gamma_+$ is the contribution from diagrams of Figs.~1a,1b
($\propto \mid T_+ \mid^2$), and $\Gamma_-$ from the diagram of
Fig.~2 ($\propto \mid T_- \mid^2$). Here we see that the width
$\Gamma(H^o \to W^+W^-Z)$ is not substantially affected by small $\xis$
parameter. Anyway, this width does not provide an experimental signature
for detecting CP-violation. $\Gamma^{MSM}$ for this decay
have also been calculated by other authors (~\cite{ll},~\cite{np}, and
references therein).

We find that the parameter $\Gamma_{fb}$ (eq.~(10)) and the related
${\cal A}_{fb}$ (eq.~(18)) are possibly relevant {\it in general}
for experimental
investigations of purely bosonic CP-violation effects. We calculated this
quantity within the minimal extension of the SM (two Higgs doublets), and
found that $\Gamma_{fb}$ may be appreciable, although in general much
smaller than $\Gamma(H^o \to W^+W^-Z)$. Interestingly enough, the final
state interactions do not represent any problem, i.e.~they do not give
any spurious (CP-conserving) contributions to $\Gamma_{fb}$. On the
other hand, the final state interactions (dominated by $W$ and $Z$-width)
are in fact crucial, together with $\xis \neq 0$, for the nonzero
CP-violation signal $\Gamma_{fb} \neq 0$. Furthermore, the CP-violation
coming from the CKM-matrix would not give any contribution to CP-violating
effects for the considered decay at the tree or 1-loop level, but possibly
only at the 2-loop level, and can be therefore safely ignored.

LHC and SSC would be able to produce Higgs with mass of several hundred
GeV (if such a Higgs exists), mostly through the gluon fusion
mechanism~\cite{gf} and the intermediate boson fusion
mechanism~\cite{ibf}. Taking the yearly estimated event rate at LHC
for the integrated luminosity to be $10^{41} cm^{-2}$ ($10^{40} cm^{-2}$
at SSC), we expect roughly $10^3$ events $H^o \to W^+W^-Z$ per year
\footnote{In ref.~\cite{ll}, the numbers $N$ of decays are given
for the minimal SM, but can be used also in the present model as
order-of-magnitude estimates.}.

Several sources of background would pose a problem for identifying
such events - particularly the direct production of $W^+W^-Z$
($p \bar p \to W^+W^-Z$) and the QCD continuum ($p \bar p \to
Z + 4 jets$). It has been argued~\cite{ll} that the background
effects of the direct production would not
be a major problem for measuring $\Gamma(H^o \to W^+W^-Z)$ at SSC for
$M_H \approx 500-600 GeV$.

However, $\Gamma_{fb}(H^o \to W^+W^-Z)$ is such a difference of the
widths for which any background
effects of the direct production that do not violate CP-symmetry
are cancelled out. To see
this, we must recall that in the ``forward-backward'' difference of
events $N_{fb}(H^o \to W^+W^-Z)$ ($\propto \Gamma_{fb}$) we make the
sum (average) over the polarizations of the incoming constituent particles
(unpolarized). For the case of direct production we choose the spin basis
$\mid S, S_z \rangle$ for polarizations of the initial $q \bar q$
(or $gg$) states
\footnote{We are allowed to take any convenient polarization basis,
since at the end we sum over all initial polarizations.}.
These initial states have well-defined CP ($CP(q \bar q) =
(-1)^{S_{q \bar q} + 1}$, $CP(gg) = (-1)^{S_{gg}}$), and hence also the
resulting directly produced $W^+W^-Z$ states would have the same
well-defined CP, provided the direct production processes themselves
do not contain appreciable CP-violating vertices. Therefore, such events
would contribute zero to $N_{fb}(W^+W^-Z)$ ($\propto \Gamma_{fb}(W^+W^-Z)$).
This argument also holds if the initial $q \bar q$ have
opposite polarizations ($S=1$).
The CKM-type CP-violation effects in the direct production could in
principle contribute to $N_{fb}$, but their effects are very small
for $q \bar q$ ($q = u,d$) and $gg$ initial states (CP-violating
asymmetries $ \stackrel{<}{\sim} {\cal O}(10^{-6})$)~\cite{np2}.

Most of the QCD continuum background may be eliminated with on-mass-shell
constraints and certain additional cuts~\cite{ll}. However, several aspects
of this problem
remain open, and this background may pose a problem for determining
$\Gamma_{fb}$.

It is possible that the CP-violating phase $\xi_{\ast}$ also enters the
general Yukawa couplings, specifically the coupling between $H^o$
and the top quark. Within the latter scenario,
D.~Chang and W.-Y.~Keung~\cite{ck} have recently investigated CP-violating
asymmetries for the decays $H^o \to W^+W^-, t \bar t$, where the source of the
asymmetries is the imaginary part ($\propto \xi_{\ast}$) of the Yukawa
coupling of $H^o$ to the top quark. They concluded that such asymmetries
can be measurable in future colliders such as SSC or LHC. These asymmetries
were ${\cal A}(H^o \to W^+W^-) = (N(W^+_LW^-_L)-N(W^+_RW^-_R))/N(W^+W^-)$
and ${\cal A}(H^o \to t \bar t) = (N(t_L \bar t_L)-N(t_R \bar t_R))/
N(t \bar t)$, where $L$ and $R$ stand for helicities $-1$, $+1$, respectively.
They found out that ${\cal A}(H^o \to W^+W^-) \stackrel{<}{\sim}
{\cal O}(10^{-3})$ and
${\cal A}(H^o \to t \bar t) \stackrel{<}{\sim} {\cal O}(10^{-1})$, for
$m_t \approx 150 GeV$. Furthermore, the branching ratios for these decays
are larger by one order of magnitude than those for $H^o \to W^+W^-Z$.
For the latter decay, ${\cal A}_{fb}$ in most cases does not exceed
${\cal O}(10^{-3})$, according to eq.~(18) and Table 1. On
the other hand, the helicity asymmetries ${\cal A}(H^o \to W^+W^-)$,
${\cal A}(H^o \to t \bar t)$
cannot be measured directly, but have to be decoded from the asymmetry of
the energy distributions of the resulting final leptons. It appears that
roughly one order of magnitude is lost due to this decoding, i.e.
${\cal A}_{l^+l^-} = ( \langle E_{l^-} \rangle - \langle E_{l^+}
\rangle )/ \langle E_{l^+} \rangle$ are about one order of magnitude
smaller than
the corresponding ${\cal A}(H^o \to W^+W^-)$,
${\cal A}(H^o \to t \bar t)$
(the decoding for $H^o \to ZZ$ is even harder). Hence, comparing
the results of ref.~\cite{ck} with the results of the present paper, we
conclude that the measurability of ${\cal A}(H^o \to W^+W^-)$ (and the
problems connected with it) is comparable to the measurability of
${\cal A}_{fb}(H^o \to W^+W^-Z)$, while the measurement of
${\cal A}(H^o \to t \bar t)$ (for $m_t \approx 150 GeV$) clearly appears
to be more promising.

One major problem in measuring $\Gamma_{fb}$ (or ${\cal A}_{fb}$)
would be a somewhat
low production rate of heavy Higgs at SSC and LHC ($N \sim 10^3$ decays
$H^o \to W^+W^-Z$ per year).
Since ${\cal A}_{fb}(H^o \to W^+W^-Z)$ ($=(N_{forw.}-N_{backw.})/N$),
in the framework discussed here, is in most cases not exceeding
${\cal O}(10^{-3})$ (eq.~(18) and Table 1), many
years of measurements would be needed to obtain possibly statistically
significant effects. Nonetheless, we believe that the proposed quantity
$\Gamma_{fb}$ (or ${\cal A}_{fb}$), or related quantities,
may eventually become relevant for
experimental tests of CP-violation of the purely bosonic sector.
Furthermore, the $\Gamma_{fb}$ and ${\cal A}_{fb}$
parameters should be investigated
numerically also for the case of larger $\xis$ parameter of CP-violation,
and they may be substantially larger in this case.

\vspace{0.4cm}

\noindent {\Large \bf Acknowledgement}

\vspace{0.4cm}

The author would like to thank K.J.~Abraham, M.~Nowakowski, E.A.~Paschos,
A.~Pilaftsis and Y.-L.~Wu for helpful discussions, particularly concerning
the possible CKM-type of CP-violation effects.
The author wishes to thank the Deutsche Forschungsgemeinschaft (DFG) and
the CEC Science project No. SC1-CT91-0729 for financial support during
the progress of this work.

\newpage

\noindent {\Large\bf Tables}

\vspace{1.cm}
\begin{table}[h]
 \begin{center}
 Table 1\\
\vspace{0.5cm}
  \begin{tabular}{|c||c|c|c||c|c|c|}  \hline
$M_{H^o}$ & $\Delta^{(1)}$ & $\Delta^{(2)}$ &  $\Delta^{(3)}$
& $\rho^{(1)}\cdot10^3$ & $\rho^{(2)}\cdot10^3$ & $\rho^{(3)}\cdot10^3$  \\
\hline
 $300$  &  $0.26$  &  $0.38$  &  $0.56$  &
$0.98$ & $1.44$ & $2.10$  \\ \hline
 $400$  &  $9.8$   &  $13.0$  &  $17.2$  &
$1.49$ & $2.00$ & $2.65$  \\ \hline
 $500$  &  $40.5$   &  $52.0$  &  $67.0$  &
$1.16$ & $1.50$ & $1.92$ \\ \hline
 $600$  &  $101.$   &  $124.$  &  $155.$  &
$0.91$ & $1.12$ & $1.40$ \\ \hline
 $700$ & $196.$ & $236.$ & $300.$ &
$0.73$ & $0.88$ & $1.10$ \\ \hline
 $800$ & $330.$ & $400.$ & $492.$ &
$0.61$ & $0.74$ & $0.91$ \\ \hline
  \end{tabular}
\vspace{0.5cm}
\caption{$\Delta$ widths (in $keV$) and $\rho$ numbers
(eqs.~(17),(18)) for various values
(in $GeV$) of the heavy Higgs mass $M_{H^o}$ ($=M_1$). The superscripts
$(1),(2),(3)$ denote these values for the cases when the masses
$(M_2, M_3)$ of the other two physical scalars (in $GeV$) are $(0,0)$,
$(100,0)$ or $(0,100)$, and $(100,100)$, respectively.}
 \end{center}
\end{table}

\begin{table}[h]
\vspace{0.4cm}
 \begin{center}
Table 2\\
\vspace{0.5cm}
  \begin{tabular}{|c||c|c|c||c|}  \hline
$M_{H^o}$  & $G^{(1)}$ & $G^{(2)}$ & $G^{(3)}$ & $\Gamma^{MSM}(H^o \to WWZ)$ \\
\hline
 $300$ & $2.3\cdot10^{-4}$ & $5.0\cdot10^{-4}$ & $1.12\cdot10^{-3}$ &
$1.16\cdot10^{-3}$ \\ \hline
 $400$ & $5.0\cdot10^{-3}$ & $9.4\cdot10^{-3}$ & $1.84\cdot10^{-2}$ &
$2.88\cdot10^{-2}$ \\ \hline
 $500$ & $2.04\cdot10^{-2}$ & $3.6\cdot10^{-2}$ & $6.5\cdot10^{-2}$ &
$1.54\cdot10^{-1}$ \\ \hline
 $600$ & $5.15\cdot10^{-2}$ & $8.6\cdot10^{-2}$ & $0.152$ &
$0.490$ \\ \hline
 $700$ & $0.103$ & $0.167$ & $0.287$ &
$1.18$ \\ \hline
 $800$ & $0.179$ & $0.285$ & $0.480$ &
$2.39$ \\ \hline
  \end{tabular}
 \caption{$G$ widths (eq.~(20)), and the decay width $\Gamma^{MSM}$
of the minimal SM, for various
values of the Higgs mass. All values are in $GeV$. The superscripts of $G$
have the same meaning as those in Table 1.}
 \end{center}
\end{table}

\clearpage

\end{document}